\documentclass[aps,prl,twocolumn,showpacs]{revtex4}
\usepackage{graphics,graphicx}

\usepackage{color}
\usepackage{wrapfig}
\usepackage{enumerate}
\usepackage{amssymb,amsmath}
\usepackage{bm}
\usepackage[normalem]{ulem} 

\begin{document}
\title{Columnar antiferromagnetic order and spin supersolid phase on the extended Shastry-Sutherland lattice}

\author{Keola~Wierschem and Pinaki~Sengupta}
\affiliation{School of Physical and Mathematical Sciences, Nanyang Technological University, 21 Nanyang Link, Singapore 637371}

\date{\today}

\begin{abstract}
We use large scale quantum Monte Carlo simulations to study an extended version of the canonical Shastry-Sutherland model -- including additional interactions and exchange anisotropy -- over a wide range of interaction parameters and an applied magnetic field. The model is  appropriate for describing the low energy properties of some members of the rare earth tetraborides. Working in the limit of large Ising-like exchange anisotropy, we demonstrate the stabilization of columnar antiferromagnetic order in the ground state at zero field and an extended magnetization plateau at $1/2$ the saturation magnetization in the presence of an applied longitudinal magnetic field -- qualitatively similar to experimentally observed low-temperature phases in ErB$_4$. Our results show that for an optimal range of exchange parameters, a spin supersolid ground state is realized over a finite range of applied field between the columnar antiferromagnetic phase and the magnetization plateau. The full momentum dependence of the longitudinal and transverse components of the static structure factor is calculated in the spin supersolid phase to demonstrate the simultaneous existence of diagonal and off-diagonal long-range order. Our results will provide crucial guidance in designing further experiments to search for the interesting spin supersolid phase in ErB$_4$.
\end{abstract}
\maketitle


Ever since Penrose and Onsager~\cite{Penrose1956} speculated on the possible coexistence of diagonal and off-diagonal long range order, supersolid order has been of broad interest within the physics community. Although the initial inspiration for this exotic type of ordering originated from the consideration of the remarkable properties of solid helium, the realization of a supersolid phase of helium remains controversial.~\footnote{For a recent review, see Ref.~\cite{Boninsegni2012}.} On the other hand, theoretical studies of several models of lattice bosons with competing interactions have conclusively established the presence of supersolid phases over extended parameter regimes.~\cite{Sengupta2005,Boninsegni2005,Heidarian2005,Melko2005,Wessel2005,Wang2009,Jiang2009} In this case, the discreteness of the lattice simplifies the process of forming a supersolid, and the bosonic models can potentially be experimentally realized with cold atoms in optical lattices. Concurrently, the realization of BEC of magnons and other novel bosonic phases in quantum magnets provides an alternative route to this elusive state of matter via the spin analog of the supersolid phase.~\cite{Matsuda1970,Liu1973,Ng2006,Laflorencie2007,Sengupta2007a,Sengupta2007b,Chen2010,Albuquerque2011} Yet to date, no experimental system has been found that displays unambiguous signs of supersolid ordering. Hence, the identification of realistic systems with supersolid order is a valuable step towards the ultimate realization of this elusive phase.

A key element in the stabilization of a supersolid ground state in many models is geometric frustration,~\cite{Boninsegni2005,Heidarian2005,Melko2005,Wessel2005,Wang2009,Jiang2009,Liu1973,Sengupta2007a,Chen2010,Albuquerque2011,Chen2008} which makes frustrated quantum magnets the natural place to look for the spin supersolid phase. The rare-earth tetraborides (RB$_4$) are a promising class of materials for studying the effects of geometric frustration in interacting spin systems. RB$_4$ compounds consist of weakly coupled layers of magnetic moment carrying R$^{3+}$ ions arranged in a distorted square lattice geometry with additional bonds along the orthogonal diagonals of alternate plaquettes~\cite{Yin2008} -- a pattern that is topologically equivalent to the Shastry-Sutherland lattice (SSL).~\cite{Shastry1981} Several members of this family have been observed to exhibit magnetization plateaus at low temperatures. For example, both ErB$_4$ and TmB$_4$ exhibit an extended magnetization plateau at $m/m_{sat}=1/2$,~\cite{Michimura2006,Yoshii2006} where $m$ and $m_{sat}$ are the uniform and saturation magnetization per site,
respectively. However, additional fractional magnetization plateaus have been observed in TmB$_4$ that are not seen in ErB$_4$. Further, recent neutron scattering experiments have determined the low temperature zero-field magnetic structure of TmB$_4$ to be collinear and antiferromagnetic (AFM) with a ${\bf Q}=(\pi,\pi)$ ordering of the local moments,~\cite{Siemensmeyer2008,Michimura2009} i.e. a staggered AFM (SAFM) state. In contrast, it is well-documented that the magnetic order of the zero-field ground state in ErB$_4$ is a columnar AFM (CAFM) pattern with ${\bf Q}=(\pi,0)$ or ${\bf Q}=(0,\pi)$.~\cite{Schafer1976,Michimura2006}

Previous studies have shown that the low temperature magnetic properties of TmB$_4$ can be explained by an effective low energy model obtained by extending the canonical Shastry-Sutherland model (SSM)~\cite{Shastry1981} to incorporate ferromagnetic (FM) transverse exchange with Ising-like anisotropy and additional long range interactions.~\cite{Suzuki2009,Suzuki2010} The additional interactions are instrumental in stabilizing an extended magnetization plateau at $m/m_{sat}=1/2$, while at the same time eliminating the magnetization plateau at $m/m_{sat}=1/3$ that is ubiquitous to the canonical SSM in the Ising limit.~\cite{Meng2008,Liu2009} Similar generalizations are expected to capture the low-energy properties of other members of the RB$_4$ family. Thus, it is useful to extend the results of this effective model to include magnetization processes that begin from the zero-field CAFM state.

In this Letter, we determine the parameter regime of the effective model for RB$_4$ for which CAFM ordering is stabilized in the absence of an external field. We present large scale quantum Monte Carlo (QMC) simulations of the magnetization process in this regime and demonstrate the appearance of a field-induced spin supersolid phase at magnetizations below $m/m_{sat}=1/2$. Arguing that similar factors may drive the stabilization of columnar order in ErB$_4$, we identify ErB$_4$ as a promising candidate for the observation of supersolid order.

The R$^{3+}$ ions typically carry a large magnetic moment, e.g., $J=6$ for Tm$^{3+}$ and $J=15/2$ for Er$^{3+}$. In many of these compounds, a strong crystal electric field introduces a single-ion anisotropy $D$ that splits the local Hilbert space at each site into degenerate doublets. Successive doublets are separated by energy gaps $\sim D$. To a first approximation, then, the low-energy magnetic interactions are captured by an Ising model comprised of the maximal-$J$ doublet. Higher order processes lead to weak FM (AFM) transverse exchange interactions between neighboring sites for integer (half-odd integer) values of $J$, and thus the $S=1/2$ XXZ model with Ising-like exchange anisotropy becomes the preferred model to study the low-energy magnetic properties of ErB$_4$ and TmB$_4$, both of which possess a large easy-axis single-ion anisotropy.~\cite{Matas2010}

\begin{figure}[tp]
\includegraphics[clip,trim=6cm 11cm 4cm 1cm,width=\linewidth]{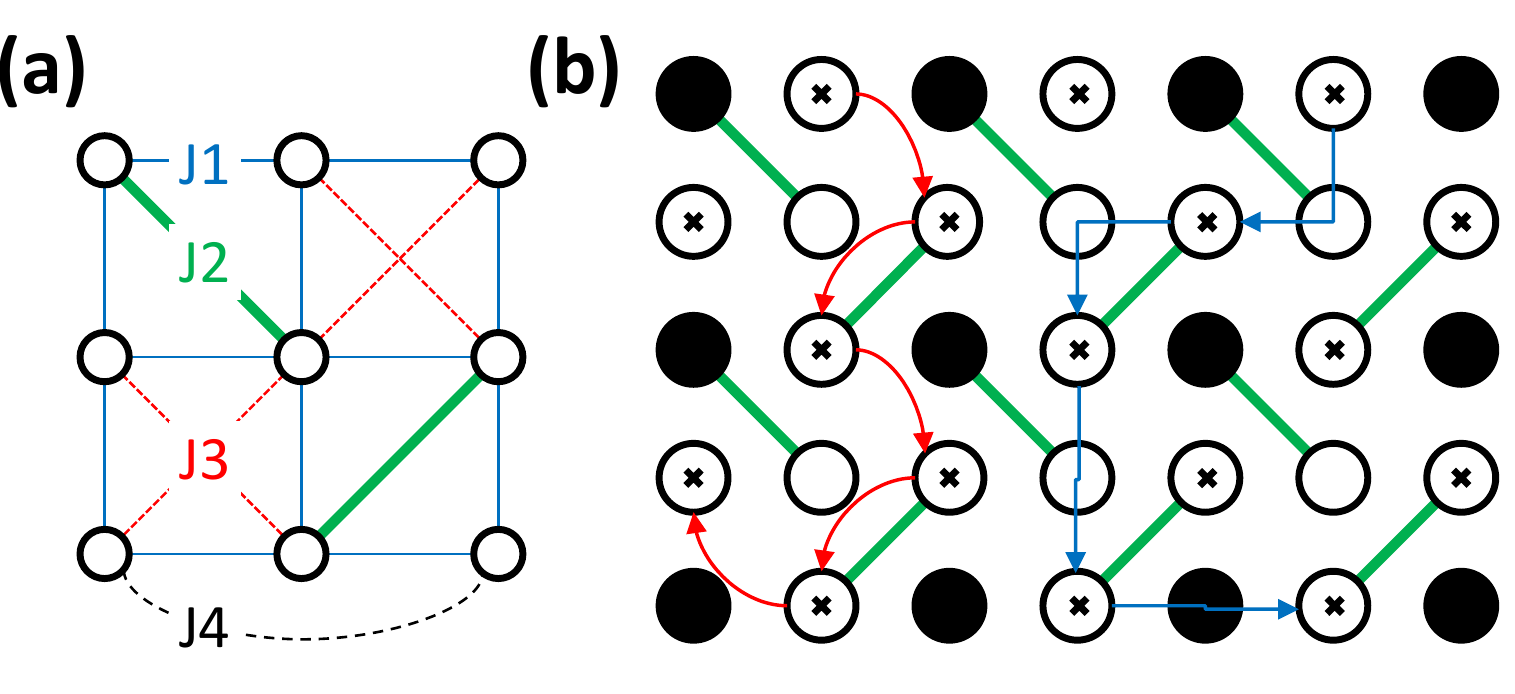}
\caption{(Color online) (a) The bonds $J_1$, $J_2$, $J_3$, and $J_4$ are illustrated as solid blue, thick green, dotted red, and dashed black lines, respectively. For clarity, only a single $J_4$ bond is shown. (b) First and second order hopping processes in the 1/2 plateau (HP) phase. White (black) circles represent up (down) spins, while an ``x'' marks spins that may be flipped with minimal energy cost. Here, the bonds $J_1$, $J_3$, and $J_4$ are omitted for clarity.}
\label{fig1}
\end{figure}

The above model can be described by the Hamiltonian
\begin{eqnarray}
{\cal H} &=& \sum_{\alpha=1}^4 \sum_{\langle ij\rangle_{\alpha}} 
\left[ -\left|J_{\alpha}\Delta\right| (S^x_i  S^x_j + S^y_i  S^y_j) + J_{\alpha} S^z_i S^z_j \right] \nonumber \\
& & -h_z\sum_i S_i^z ,
\label{eq:H}
\end{eqnarray}
where the summation is over all bond types $\alpha$ of strength $J_{\alpha}$ and their associated bonds $\langle ij\rangle_{\alpha}$ [Fig.~\ref{fig1}(a)]. We work in the experimentally relevant limit of strong Ising-like exchange anisotropy, $\Delta\ll1$, and consider the effect of an external magnetic field $h_z$.  Since the RB$_4$ crystal structure is such that the bonds of the canonical SSL are expected to be of roughly equal strength, we set $J_1=J_2=1$, and all parameters are given in these units.~\footnote{Note that while we have set $J_1$ identically equal to $J_2$, similar results are expected for any $J_1/J_2$ ratio near unity. We have explicitly verified that this is the case from $J_2=0.9J_1$ to $J_2=1.1J_1$.} In addition to the canonical interactions $J_1$ and $J_2$ of the SSM, we include the interactions $J_3$ and $J_4$. As shown below, $J_3$ is the primary interaction driving the CAFM state. The role of $J_4$ is to stabilize the magnetic states. Extensive simulations (not shown here) in the $J_3-J_4$ parameter space have shown that $J_4<0$ generates a sequence of plateaus that qualitatively matches the experimental data for ErB$_4$. On the other hand, even a small $J_4>0$ leads to a different sequence of plateaus. Interestingly, $J_4<0$ was  found to be instrumental in explaining the magnetic properties of TmB$_4$.~\cite{Suzuki2009,Suzuki2010} Such a long range interaction is intuitively expected to arise from RKKY interactions in the RB$_4$ compounds mediated by the itinerant electrons.~\cite{Siemensmeyer2008}

In this work, we restrict ourselves to the case of FM transverse exchange interactions, to avoid the negative sign problem in QMC simulation. This is in contradiction to the effective transverse exchange in ErB$_4$. However, it was shown recently~\cite{Wang2009,Jiang2009} that in the Ising limit, an AFM transverse exchange maps on to a FM transverse exchange in the triangular lattice XXZ model. As demonstrated later, similar arguments also hold for the present work.

To characterize the different phases of ${\cal H}$, longitudinal and transverse components of the static structure factor are defined as
\begin{eqnarray} \label{eq:eq2}
S^{+-}(\bm k) &=& \frac{1}{N}\sum_{i,j}
	\langle S^+_i S^-_j + S^-_i S^+_j \rangle 
	e^{i{\bm k}\cdot({\bm r}_i-{\bm r}_j)}, \nonumber \\
S^{zz}(\bm k) &=& \frac{1}{N}\sum_{i,j}
	\langle S^z_i S^z_j \rangle 
	e^{i{\bm k}\cdot({\bm r}_i-{\bm r}_j)}.
\end{eqnarray}
In the $S^z$ basis, these serve as measures of off-diagonal and diagonal order, respectively. As the uniform magnetization per site is simply $m\equiv\langle m_z\rangle=\sum_j \langle S^z_j \rangle/N$, here we note that $S^{zz}(0,0)=N\langle m^2_z\rangle$. Further, we define staggered and columnar magnetizations as $m^2_s=S^{zz}(\pi,\pi)/N$ and $m^2_c=\left[S^{zz}(0,\pi)+S^{zz}(\pi,0)\right]/N$, respectively.

QMC simulations are performed using the stochastic series expansion algorithm with directed loop updates.~\cite{Syljuasen2002} The correlations $\langle S^+_i S^-_j + S^-_i S^+_j \rangle$ and $\langle S^z_i S^z_j \rangle$ can be straightforwardly computed within this method~\cite{Sandvik1991,Dorneich2001}. A useful observable in characterizing the ground state phases in numerical simulations is the spin stiffness, $\rho_s$, defined as the response to a twist in the boundary conditions.~\cite{Sandvik1997} In simulations that sample multiple winding number sectors, the evaluation of the stiffness simplifies to calculating the winding number of the world lines, leading to the expression $\rho_s=\left(w_x^2+w_y^2\right)/2\beta$ in two dimensions,~\cite{Pollock1987} where $w_x$ and $w_y$ are the winding numbers in the $x$ and $y$ directions.

\begin{figure}[t]
\includegraphics[clip,trim=0cm 1cm 0cm 2cm,width=\linewidth]{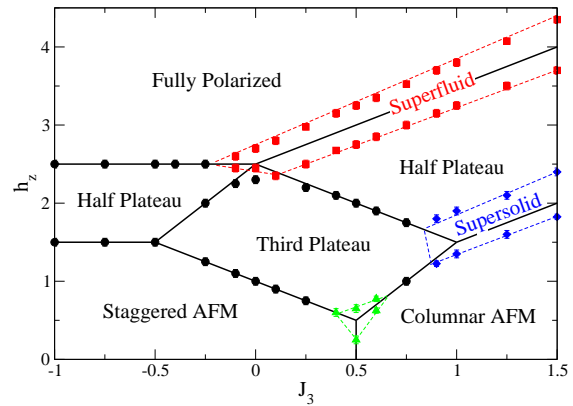}
\caption{(Color online) Phase diagram in the Ising limit (solid black lines) along with modifications for the XXZ model at Ising-like anisotropy $\Delta=0.10$ (data points and colorful dashed lines). The saturation field from the superfluid phase is $h_{sat}=\left(5/2+J_3\right)\left(1+\Delta\right)$, while the remaining quantum phase boundaries are drawn as guides to the eye.}
\label{fig2}
\end{figure}

We begin by mapping out a phase diagram for ${\cal H}$ with $J_4=0$ to show that $J_3$ is sufficient to stabilize the CAFM ground state. In the Ising limit, we can construct the ground states by hand. By comparing their respective energies, the $h_z-J_3$ phase diagram is determined (see Fig.~\ref{fig2}). This provides a backdrop of classically ordered magnetic states upon which to study the effect of quantum fluctuations. The zero-field ground state is SAFM for $J_3<1/2$ and CAFM for $J_3>1/2$. For small $|J_3|$ there is a field-induced magnetic phase transition to a third plateau (TP) with $m/m_{sat}=1/3$. With increasing field, this state further evolves into a half plateau (HP) with $m/m_{sat}=1/2$, except at $J_3=0$, where there is a direct transition to the fully polarized (FP) state ($m/m_{sat}=1$).~\cite{Chang2009} At large $|J_3|$ the magnetization process skips the TP, with a direct transition from the zero-field ground states to the HP. Notice that two distinct magnetic structures are possible in the HP state.~\cite{Dublenych2012} For $J_3<0$ the HP consists of diagonal stripes with staggered order, while for $J_3>0$ the HP consists of alternating FM and AFM stripes with columnar order (the AFM stripes are free to align or anti-align, resulting zero net staggered order). Thus, the two HP states mirror the order of the underlying zero-field ground states (SAFM and CAFM, respectively).

Turning to the case of finite $\Delta=0.10$, QMC simulations are performed to determine the ground state quantum phase diagram. In Fig.~\ref{fig2}, the numerically determined phase boundaries are marked by data points, while three emergent phases with transverse ferromagnetic order (i.e. superfluid) are delineated by dashed lines. The first emergent phase (red squares) occurs primarily for $J_3>0$, where a superfluid (SF) phase opens up between the HP and FP phases. A similar phase develops between the CAFM and HP phases for $J_3\ge0.9$. This second emergent phase (blue diamonds) is actually a spin supersolid (SS) phase, possessing both superfluid and columnar AFM order. A third emergent phase (green triangles) with superfluid ordering occurs at the border of the SAFM, CAFM, and TP phases.

\begin{figure}[htp]
\includegraphics[clip,trim=0cm 1cm 0cm 1cm,width=\linewidth]{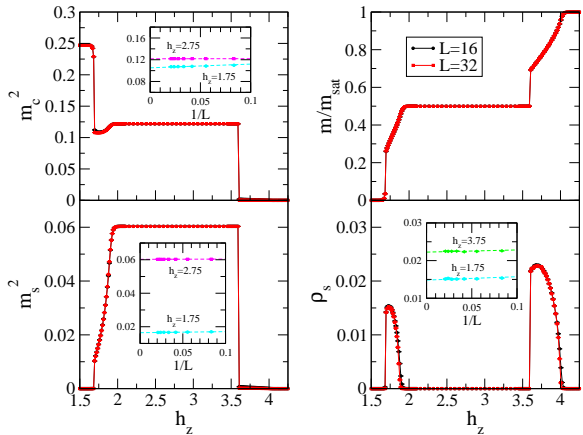}
\caption{(Color online) Columnar ($m_c^2$), staggered ($m_s^2$), and uniform ($m$) magnetization, along with spin stiffness ($\rho_s$), for the complete magnetization process at $\Delta=0.10$, $J_{3}=1.25$, $J_{4}=-0.05$, and inverse temperature $\beta=64$. The sequence of phases with increasing magnetic field is columnar AFM, supersolid, half plateau, superfluid, and fully polarized. Insets: Finite-size scaling of $m_c^2$, $m_s^2$, and $\rho_s$. Dashed lines are fits to the form $a+b/L$.}
\label{fig3}
\end{figure}

Focusing on the experimentally relevant parameter regime for ErB$_4$ ($J_3>1$), the introduction of $J_4<0$ is desirable to stabilize the magnetization process (as discussed earlier, long range interactions such as $J_4$ are expected to be non-negligible due to RKKY effects, and $J_4>0$ leads to a qualitatively different sequence of magnetization plateaus). The main result of $J_4<0$ is to stabilize the HP such that the AFM stripes align, as in Fig.~\ref{fig1}(b). In Fig.~\ref{fig3} we show the magnetization process for $\Delta=0.10$, $J_3=1.25$, and $J_4=-0.05$. Here, the zero-field ground state is the CAFM phase with $m_c^2$ near its maximal value of $0.25$. As the magnetic field is increased, there is a field-induced first-order phase transition from the CAFM phase to the SS phase, as evidenced by discontinuous jumps in the magnetic observables. In the SS phase, $m_c^2$ is reduced, yet remains finite, and $m_s^2$ becomes non-zero, even as a non-zero spin stiffness develops. These are signs of supersolid order, which we carefully consider in the next paragraph. After a continuous transition to the HP state, diagonal order remains while superfluid order is eliminated. Next, a first-order phase transition (with accompanying jumps in the magnetic observables) eliminates all remaining diagonal magnetic order, and the HP gives way to a SF phase with off-diagonal ordering. Finally, there is a continuous phase transition from the SF to the FP state. Although both continuous phase transitions (SS-HP and SF-FP) are expected to belong to the BEC universality class, their numeric confirmation is outside the scope of the present work.

\begin{figure}[t]
\vspace*{0cm}
\includegraphics[clip,trim=1cm 3cm 1cm 4cm,width=\linewidth]{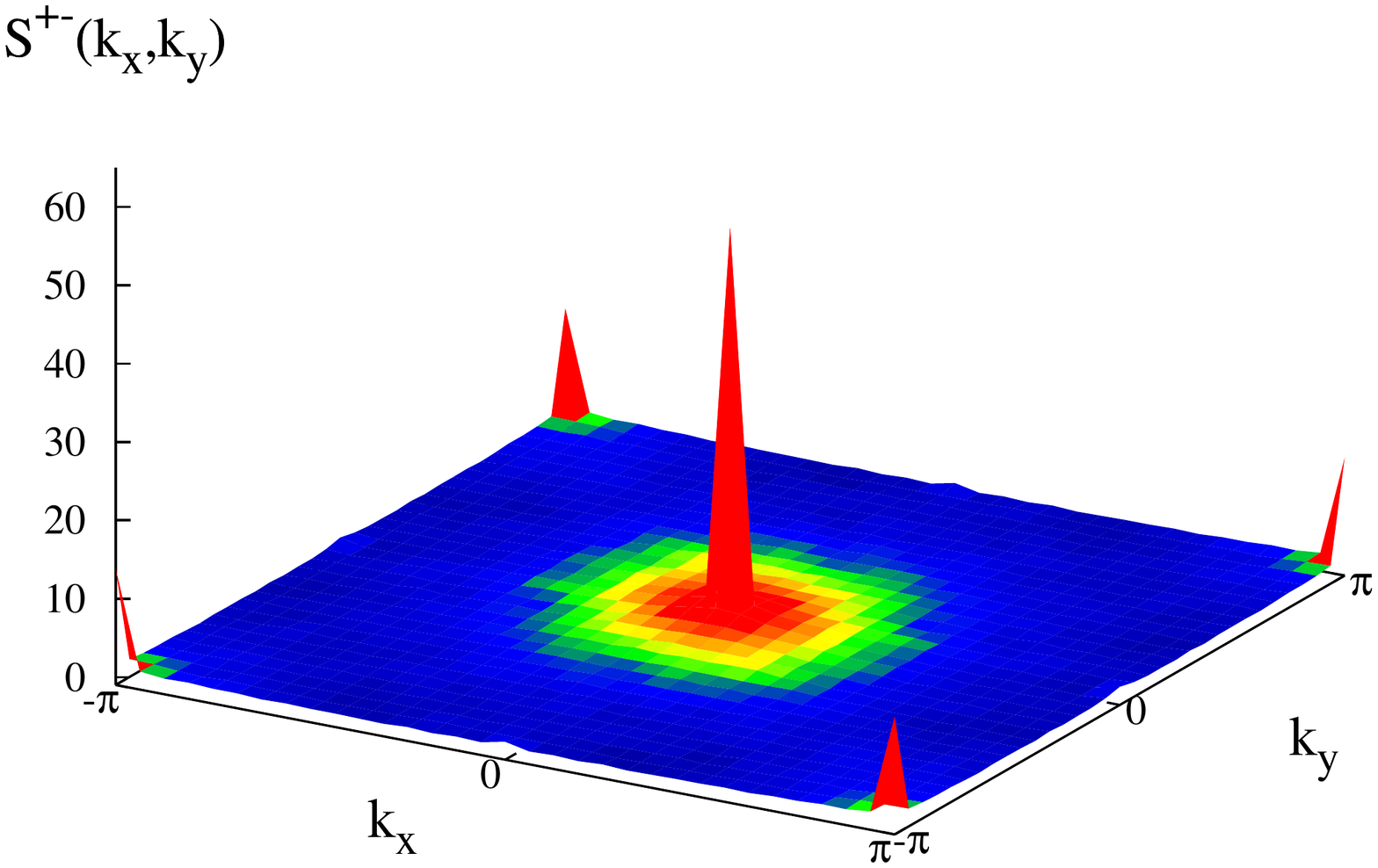}
\includegraphics[clip,trim=1cm 3cm 1cm 4cm,width=\linewidth]{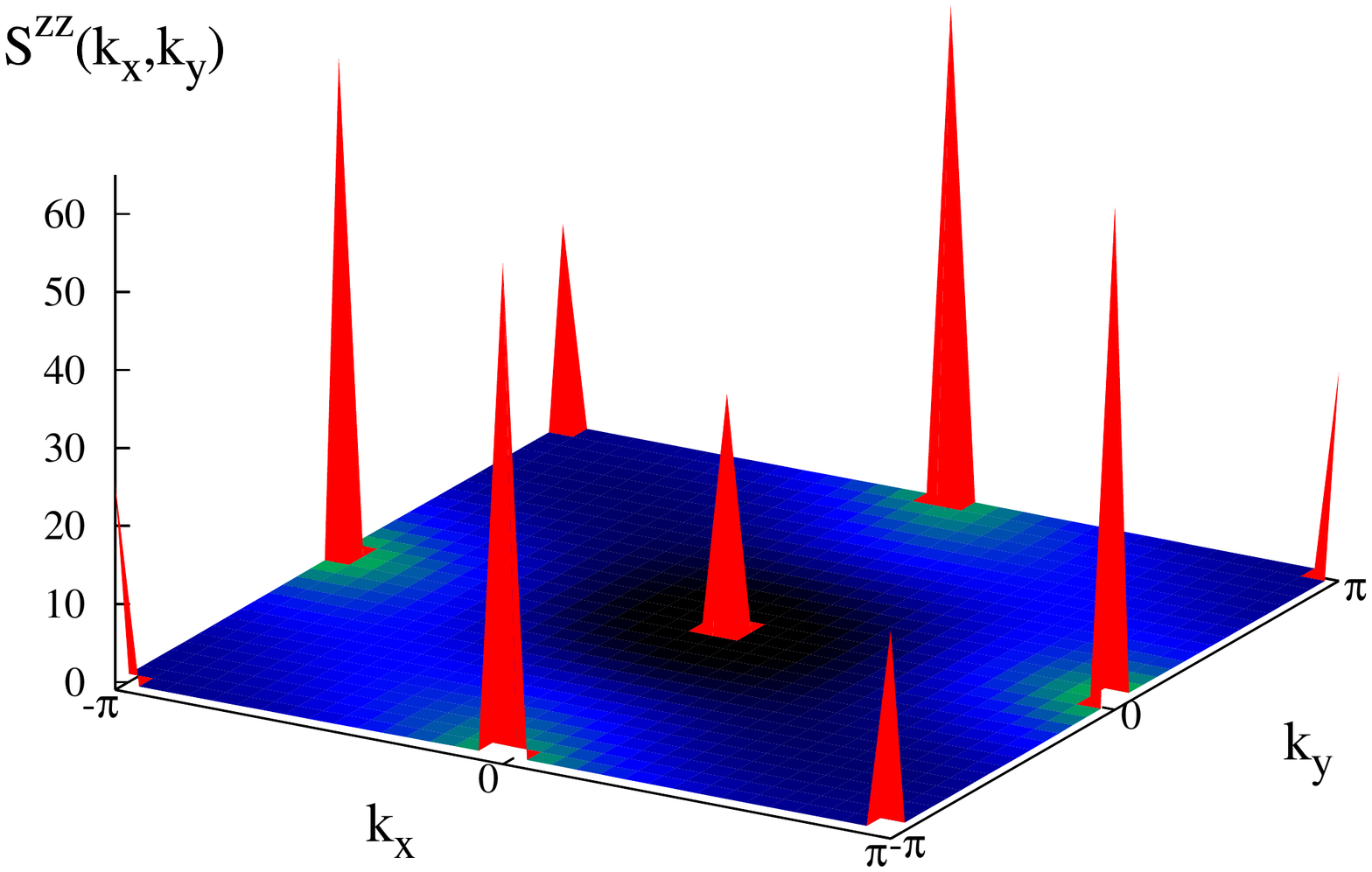}
\caption{(Color online) Transverse and longitudinal structure factors in the spin supersolid phase, with $S^{+-}(\bm k)$ and $S^{zz}(\bm k)$ as defined in Eq.~\ref{eq:eq2}. Data are taken from a 32x32 lattice at inverse temperature $\beta=100$ with $\Delta=0.10$, $h_z=1.75$, $J_{3}=1.25$, and $J_{4}=-0.05$.}
\label{fig4}
\end{figure}

A supersolid is defined by the simultaneous occurrence of diagonal and off-diagonal long-range order. To  confirm  that the present spin supersolid phase satisfies this property, we calculate the full static structure factors of the transverse and longitudinal degrees of spin freedom ($S^{+-}$ and $S^{zz}$, respectively). As seen in Fig.~\ref{fig4}, it is clear that both structure factors possess ordering. The transverse ordering is primarily at ${\bm k}=(0,0)$, with a secondary peak at ${\bf k}=(\pi,\pi)$ due to modulation by the diagonal order. The peaks in $S^{zz}$ simply reflect the longitudinal magnetic ordering ($m_c^2\ne0$ and $m_s^2\ne0$) and finite magnetization ($m\ne0$) of the SS phase.

The mechanism for superflow in the SS phase can further explain the observed features of the transverse and longitudinal structure factors. Since the SS phase occurs at magnetizations just below half saturation, the mechanism for supersolid formation can be determined by considering the hopping processes available when the HP state is doped with down spins. The lowest potential energy doping sites in the HP phase are marked by an ``x'' in Fig.~\ref{fig1}(b). It is clear to see that down spins at these sites can easily delocalize by first order processes. Furthermore, these sites form an unfrustrated sublattice, naturally explaining the secondary peak in $S^{+-}$ at ${\bf k}=(\pi,\pi)$. Another consequence of this unfrustrated sublattice of first order hopping processes occurs as we approach the Ising limit ($\Delta\rightarrow0$), where higher order hopping processes are suppressed. Here, models with AFM and FM transverse exchange become essentially equivalent, and QMC results with FM transverse exchange can be mapped to the case of AFM transverse exchange. This can also be seen by projecting ${\cal H}$ onto the Ising ground state manifold,~\cite{Jiang2009} whereby higher-order processes are explicitly forbidden. We have confirmed this equivalence through exact calculations for systems up to $L=6$, and see no difference in the magnetic properties for AFM and FM transverse exchange in this limit.

The model Hamiltonian ${\cal H}$ can be mapped onto a system of hardcore bosons with nearest-neighbor hopping $t=-\left|J_\alpha\Delta\right|/2$ and repulsion $V=J_\alpha$. For such a system on the triangular lattice it has been shown~\cite{Boninsegni2005,Heidarian2005,Melko2005,Wessel2005} that a supersolid phase is stabilized for weak hopping $t\ll V$ or $\Delta\ll 1$ (a supersolid phase is also found for frustrated hopping $t>0$~\cite{Wang2009,Jiang2009}). Similarly, a square lattice model of hardcore bosons with next-nearest-neighbor repulsion is known to possess both columnar AFM and supersolid phases.~\cite{Chen2008} While no such supersolid state has been found for hardcore bosons on the canonical SSL,~\cite{Gan2011} we have demonstrated the formation of a supersolid phase on the extended SSL.

Let us return to the magnetic properties of ErB$_4$, which has a zero-field CAFM ground state and a single plateau at $m/m_{sat}=1/2$. We have shown that such properties are captured by an extended SSM with $J_3>1$ and $J_4<0$. The numerically determined magnetization process in this regime compares well to experimental observations.~\cite{Michimura2006} This agreement is not only qualitative, but also quantitative: the relative extent of the HP state is $\Delta h_{1/2}/h_{sat}\sim(3.5-2)/4$ or ~38\% in our model vs. $\Delta h_{1/2}/h_{sat}\sim(4-2)/5$ or ~40\% in experiment~\cite{Michimura2006}. The close agreement of the two magnetization processes leads us to speculate that the spin supersolid phase we have demonstrated in the model may also be present in the material. Although no estimates are known for the exchange parameters in ErB$_4$ (aside from the geometrically justified approximation of $J_1\approx J_2$), considering the electronic structure of RB$_4$ compounds,~\cite{Yin2008} it is plausible to expect $J_3>1$. We hope that our results will encourage further experimental studies of this compound. Finally, we note that superfluid ordering in the effective spin model corresponds to {\em ferronematic} order in the total angular momentum of the material.~\cite{Wierschem2012}

\begin{acknowledgments}
It is a pleasure to acknowledge fruitful discussions with C.~D.~Batista.
This research used resources of the National Energy Research Scientific Computing Center, which is supported by the Office of Science of the U.S. Department of Energy under Contract No. DE-AC02-05CH11231. 
\end{acknowledgments}

\bibliographystyle{apsrev}
\bibliography{ref}

\end{document}